\documentclass[aps,pra,onecolumn,superscriptaddress,nofootinbib]{revtex4-2}

\usepackage{amsmath,amssymb,bm}
\usepackage{xcolor}
\usepackage{hyperref}
\hypersetup{pdftitle={Cohesion accounting of complementarity in path--polarization interferometry},pdfauthor={Jose J. Gil}}

\begin{document}

\title{Cohesion accounting of complementarity in path--polarization interferometry}

\author{Jos\'e J. Gil}
\email{ppgil@unizar.es}
\affiliation{Honorary Collaborator, Department of Applied Physics, University of Cantabria, Av. Los Castros 48, 39005 Santander, Spain}

\date{\today}

\begin{abstract}
Two-path complementarity in polarized interferometric fields is reconsidered by retaining the complete path--polarization density matrix instead of reducing the description to the path degree of freedom from the outset. The familiar relation connecting the Cartesian visibility components, path predictability, and reduced-state mixedness is recovered as a marginal consequence of the reduced path state and is not interpreted as a new complementarity law. Attention is focused instead on the full path--polarization description in a real reference basis adapted to the path and linear-polarization degrees of freedom. Within this framework, the normalized purity separates naturally into path, polarization, and path--polarization-correlation contributions, while the antisymmetric sector provides a sector-resolved measure of cohesion. The resulting decomposition identifies which parts of the complete state store phase-sensitive interferometric coherence and which contributions are removed when polarization is traced out. The formalism therefore provides a sector-resolved accounting of complementarity within the full path--polarization state and clarifies the connection between reduced visibility loss, polarization marking, path--polarization correlations, and quantum-eraser recovery. The present article establishes the framework and associated purity decomposition; a more detailed exploration of full path--polarization complementarity and its dynamical aspects is left for future work.
\end{abstract}

\maketitle

\section{Introduction}

Wave--particle complementarity in two-path interferometry is commonly expressed through relations between fringe visibility, path predictability, distinguishability, and entanglement with a marker degree of freedom~\cite{Greenberger1988,Englert1996,JakobBergou2010,Coles2014}.  In optical realizations the marker is often the polarization degree of freedom, and the role of polarization modulation in vector-light interference has been analyzed in depth by Norrman \textit{et al.}~\cite{Norrman2017}.  Closely related geometric formulations have also connected coherence, which-path information, and entanglement through the Bloch sphere and stereographic constructions~\cite{Maleki2019}.  These works make clear that the reduced two-path state already contains a complete three-component Bloch-vector description of the scalar interference pattern.

This observation is important for the present paper.  If the joint path--polarization state is reduced to the path qubit, then writing the complex interpath coherence in terms of its real and imaginary parts gives only the Cartesian decomposition of a Bloch vector.  The corresponding identity
\begin{equation}
V_A^2+V_N^2+\mathcal P^2+\mathcal I^2=1
\label{eq:intro_reduced_identity}
\end{equation}
is exact and useful, but it is not by itself a new complementarity principle.  Here $V_A$ and $V_N$ are the in-phase and quadrature components of the reduced visibility, $\mathcal P$ is the path predictability, and $\mathcal I$ is the reduced-path mixedness.  Equation~\eqref{eq:intro_reduced_identity} is best understood as a marginal identity for the reduced path qubit.

The aim of this work is therefore more specific.  We ask where the phase-sensitive interferometric information is stored in the complete path--polarization state before the polarization degree of freedom is discarded.  This requires keeping the $4\times4$ density matrix on $\mathcal H_{\rm path}\otimes\mathcal H_{\rm pol}$ and resolving its real-symmetric and imaginary-antisymmetric sectors in a fixed real physical basis.  The construction is compatible with the intrinsic-reference-basis (IRB) framework used to separate intrinsic populations from real antisymmetric cohesion components~\cite{Gil2020Symmetry,Gil2026Classicality}, but the present paper deliberately retains the laboratory path--polarization basis.  This is essential because the objective is to identify which experimentally meaningful path, polarization, and correlation sectors store the phase-sensitive content that is lost under a polarization trace.

The paper is deliberately modest about the reduced identity but explicit about the full-state contribution.  The novelty claimed here is not the two-level Bloch geometry.  Rather, the novelty lies in retaining the complete path--polarization state and identifying how coherence, polarization, and path--polarization correlations contribute separately to the full-state purity and cohesion balance.  It is the use of the real-basis decomposition of the full path--polarization density matrix to distinguish reduced visibility, hidden path--polarization correlations, and recoverable coherence under polarization marking and erasure.  The formalism does not introduce a new complementarity principle; rather, it provides a sector-resolved accounting of complementarity within the full path--polarization state.

\section{Path--polarization state and reduced complementarity}

Let $\{|1\rangle,|2\rangle\}$ denote the path basis and $\{|H\rangle,|V\rangle\}$ a fixed pair of orthogonal linear-polarization states.  The complete state is a normalized positive Hermitian matrix
\begin{equation}
\rho\ge0,\qquad \operatorname{Tr}\rho=1,
\end{equation}
on the four-dimensional product space $\mathcal H_{\rm path}\otimes\mathcal H_{\rm pol}$, ordered as
\begin{equation}
\{|1H\rangle,|1V\rangle,|2H\rangle,|2V\rangle\}.
\end{equation}
The reduced path state is
\begin{equation}
\rho_{\rm p}=\operatorname{Tr}_{\rm pol}\rho
=\frac12\left(I+\bm s\cdot\bm\sigma\right),
\label{eq:path_reduced}
\end{equation}
where $\bm s=(s_x,s_y,s_z)$ is the Bloch vector of the path qubit.  With a fixed phase origin in the interferometer we define
\begin{equation}
V_A=s_x,
\qquad
V_N=s_y,
\qquad
Z=s_z,
\qquad
\mathcal P=|Z|.
\label{eq:path_components}
\end{equation}
Equivalently,
\begin{equation}
V_A=2\operatorname{Re}(\rho_{{\rm p},12}),
\qquad
V_N=-2\operatorname{Im}(\rho_{{\rm p},12}),
\qquad
Z=\rho_{{\rm p},11}-\rho_{{\rm p},22},
\end{equation}
up to the conventional sign chosen for $\sigma_y$.  The polarization-unresolved visibility is
\begin{equation}
V=2|\rho_{{\rm p},12}|=\sqrt{V_A^2+V_N^2}.
\end{equation}
Defining the reduced-path mixedness by
\begin{equation}
\mathcal I^2=1-|\bm s|^2=2\left(1-\operatorname{Tr}\rho_{\rm p}^2\right)=4\det\rho_{\rm p},
\label{eq:mixedness_reduced}
\end{equation}
we obtain immediately
\begin{equation}
V_A^2+V_N^2+\mathcal P^2+\mathcal I^2=1.
\label{eq:reduced_identity}
\end{equation}
This is the reduced two-level identity.  It implies the Greenberger--Yasin bound $V^2+\mathcal P^2\le1$~\cite{Greenberger1988}.  For globally pure path--polarization states, $\mathcal I$ coincides with the concurrence between path and polarization, and Eq.~\eqref{eq:reduced_identity} becomes the Jakob--Bergou complete-complementarity equality~\cite{JakobBergou2010,Wootters1998}.  For mixed global states, $\mathcal I$ measures mixedness of the reduced path state, not necessarily entanglement.

Equation~\eqref{eq:reduced_identity} will be used below only as a marginal constraint.  It is insufficient to identify how much coherence remains in the complete path--polarization state because the partial trace may suppress correlations that are still present and experimentally recoverable by polarization analysis or erasure.

\section{Real-basis decomposition, cohesion, and basis dependence}

Let the chosen path--polarization product basis be real.  In that basis every Hermitian state admits the unique decomposition
\begin{equation}
\rho=S+i\mathcal T,
\qquad
S=S^{\mathsf T}\in\mathbb R^{4\times4},
\qquad
\mathcal T=-\mathcal T^{\mathsf T}\in\mathbb R^{4\times4}.
\label{eq:S_iT}
\end{equation}
Here $\mathcal T=\Im\rho$ is the laboratory representation of the imaginary antisymmetric, or cohesion, bivector.  In the notation used in the companion classicality work, this laboratory object becomes the intrinsic cohesion matrix $N$ after the orthogonal transformation to the canonical IRB, as made explicit in Eq.~\eqref{eq:IRB_N}.  The notation $\mathcal T$ is retained here to avoid confusing this laboratory representation with the path--polarization correlation matrix introduced below.  In a generic real basis the scalar cohesion associated with this sector is
\begin{equation}
P_c^2=2\|\mathcal T\|_F^2.
\label{eq:Pc_lab_definition}
\end{equation}
This quantity is invariant under real orthogonal changes of basis.  In the canonical IRB construction, one first diagonalizes the real symmetric sector by an orthogonal matrix $Q_O$ and represents the same bivector as
\begin{equation}
N=Q_O^{\mathsf T}\mathcal T Q_O,
\qquad
P_c^2=2\|N\|_F^2=2\|\mathcal T\|_F^2.
\label{eq:IRB_N}
\end{equation}
Thus the global scalar cohesion is basis independent, whereas its decomposition into path, polarization, and path--polarization contributions is basis dependent.  This dependence is not a defect here.  It is the operational point of the present article, because path predictability, linear-polarization analysis, and phase-sensitive interference are defined with respect to a physically fixed laboratory basis.

Similarly, any Hermitian observable $O$ can be written as $O=O_S+iO_A$, with $O_S$ real symmetric and $O_A$ real antisymmetric.  Since the Frobenius product of a real symmetric and a real antisymmetric matrix vanishes, one has
\begin{equation}
\langle O\rangle=\operatorname{Tr}(\rho O)
=\operatorname{Tr}(SO_S)-\operatorname{Tr}(\mathcal T O_A).
\label{eq:sector_decoupling}
\end{equation}
Thus real-symmetric observables probe $S$, whereas imaginary-antisymmetric observables probe $\mathcal T$.  This is the precise meaning of sectoral coupling in the present formulation.  The statement is algebraic; the laboratory implementation of a given Pauli tensor still depends on the required path phase control and polarization analysis.

For a two-level reduced path state,
\begin{equation}
\rho_{\rm p}=S_{\rm p}+i\mathcal T_{\rm p},
\qquad
\mathcal T_{\rm p}=\frac{V_N}{2}\begin{pmatrix}0&-1\\1&0\end{pmatrix},
\end{equation}
with the same sign convention as Eq.~\eqref{eq:path_components}.  The cohesion convention gives, for the reduced path qubit,
\begin{equation}
P_{c,{\rm p}}^2=2\|\mathcal T_{\rm p}\|_F^2=V_N^2.
\label{eq:path_pc}
\end{equation}
Thus the antisymmetric reduced-path cohesion is the quadrature component of the interferometric visibility in the adopted phase convention.  The total visibility also contains the real-symmetric component $V_A$; therefore $P_{c,{\rm p}}$ is not the full visibility unless the phase origin is chosen so that $V_A=0$ or the coherence is purely quadrature.  Conversely, by shifting the interferometric phase origin one rotates $(V_A,V_N)$ without changing $V$.

\section{Full path--polarization balance}

The full $4\times4$ density matrix can be expanded in the standard two-qubit Pauli basis as
\begin{equation}
\rho=\frac14\left[
I\otimes I+\bm s\cdot\bm\sigma\otimes I+I\otimes\bm p\cdot\bm\sigma+
\sum_{i,j=x,y,z}C_{ij}\,\sigma_i\otimes\sigma_j
\right],
\label{eq:fano}
\end{equation}
where
\begin{equation}
 s_i=\operatorname{Tr}\rho(\sigma_i\otimes I),\qquad
 p_j=\operatorname{Tr}\rho(I\otimes\sigma_j),\qquad
 C_{ij}=\operatorname{Tr}\rho(\sigma_i\otimes\sigma_j).
\end{equation}
The vector $\bm s$ is precisely the reduced path Bloch vector in Eq.~\eqref{eq:path_reduced}; $\bm p$ is the reduced polarization Bloch vector; and $C$ is the path--polarization correlation matrix.  Orthogonality of Pauli tensors gives
\begin{equation}
\operatorname{Tr}\rho^2=\frac14\left(1+|\bm s|^2+|\bm p|^2+\|C\|_F^2\right).
\label{eq:trrho2}
\end{equation}
The normalized four-dimensional degree of purity is
\begin{equation}
P_\Delta^2=\frac{4\operatorname{Tr}\rho^2-1}{3}
=\frac{|\bm s|^2+|\bm p|^2+\|C\|_F^2}{3}.
\label{eq:Pdelta}
\end{equation}
Equation~\eqref{eq:Pdelta} is the standard two-qubit purity decomposition written in Bloch--correlation coordinates.  Its role here is not to introduce a new purity identity, but to resolve the path contribution into the experimentally meaningful quantities $V_A$, $V_N$, and $\mathcal P$, and to connect the resulting decomposition with the cohesion framework.

Using $|\bm s|^2=V_A^2+V_N^2+\mathcal P^2$, we obtain the full path--polarization balance
\begin{equation}
3P_\Delta^2=
V_A^2+V_N^2+\mathcal P^2+P_{\rm pol}^2+\|C\|_F^2,
\label{eq:full_balance}
\end{equation}
where
\begin{equation}
P_{\rm pol}^2=|\bm p|^2
\end{equation}
is the squared degree of polarization of the reduced polarization state.  Equation~\eqref{eq:full_balance} is elementary but useful: it shows exactly what is removed when the state is projected down to the path qubit.  The reduced identity keeps only $\bm s$ and absorbs all omitted path--polarization structure into the reduced mixedness $\mathcal I$.  The full identity instead resolves this omitted structure into intrinsic polarization and correlation weights.

Combining Eqs.~\eqref{eq:mixedness_reduced} and \eqref{eq:Pdelta} gives
\begin{equation}
\mathcal I^2=1-|\bm s|^2
=1-3P_\Delta^2+P_{\rm pol}^2+\|C\|_F^2.
\label{eq:I_full_relation}
\end{equation}
For a globally pure path--polarization state, $P_\Delta=1$, and hence
\begin{equation}
\mathcal I^2=P_{\rm pol}^2+\|C\|_F^2-2.
\label{eq:pure_I_full}
\end{equation}
This form is not by itself the most transparent expression for entanglement, but it makes explicit that reduced path mixedness is produced by full-state information not retained in the reduced path vector.  It is included only to illustrate how the mixedness of the reduced path state originates from full-state polarization and path--polarization-correlation contributions; no new entanglement measure is introduced here.

\section{Antisymmetric sector and cohesion accounting}

In the Pauli expansion, a tensor is imaginary antisymmetric in the chosen real product basis exactly when it contains one and only one factor $\sigma_y$.  Therefore the laboratory cohesion matrix $\mathcal T=\Im\rho$ of the full state is determined by the six coefficients
\begin{equation}
\mathcal N=\{s_y,\;p_y,\;C_{yx},\;C_{yz},\;C_{xy},\;C_{zy}\}.
\label{eq:N_coeffs}
\end{equation}
The tensor $C_{yy}\sigma_y\otimes\sigma_y$ is real symmetric and belongs to $S$, not to $\mathcal T$, because it contains two imaginary Pauli factors.  With the normalization of Eq.~\eqref{eq:fano}, orthogonality of the Pauli tensors yields
\begin{equation}
\|\mathcal T\|_F^2=\frac14\left(s_y^2+p_y^2+C_{yx}^2+C_{yz}^2+C_{xy}^2+C_{zy}^2\right).
\label{eq:Tcal_norm}
\end{equation}
Consequently,
\begin{equation}
P_c^2=2\|\mathcal T\|_F^2
=\frac12\left(s_y^2+p_y^2+C_{yx}^2+C_{yz}^2+C_{xy}^2+C_{zy}^2\right).
\label{eq:Pc_full}
\end{equation}
This is the total cohesion contained in the imaginary-antisymmetric part of the complete path--polarization state, expressed in the physical laboratory basis.  It separates operationally as
\begin{equation}
P_c^2=P_{c,{\rm path}}^{2({\rm glob})}+P_{c,{\rm pol}}^{2({\rm glob})}+P_{c,{\rm corr}}^{2({\rm glob})},
\label{eq:Pc_sectoral}
\end{equation}
with
\begin{align}
P_{c,{\rm path}}^{2({\rm glob})}&=\frac12 s_y^2=\frac12 V_N^2,\nonumber\\
P_{c,{\rm pol}}^{2({\rm glob})}&=\frac12 p_y^2,\nonumber\\
P_{c,{\rm corr}}^{2({\rm glob})}&=\frac12\left(C_{yx}^2+C_{yz}^2+C_{xy}^2+C_{zy}^2\right).
\label{eq:Pc_sectors}
\end{align}
The superscript ``glob'' emphasizes that these are contributions to the cohesion of the four-dimensional state.  They should not be confused with the two-dimensional cohesion of the reduced path state, Eq.~\eqref{eq:path_pc}.  The factor differs because the same path operator is embedded in a larger tensor-product space.

The quantity $p_y$ is the circular-polarization Stokes component relative to the chosen linear $H/V$ reference axes.  Its appearance in Eq.~\eqref{eq:Pc_full} does not imply path interference by itself.  Rather, it quantifies phase-sensitive structure already present in the polarization subsystem.  Whether such structure can be converted into observable path interference depends on the available path--polarization correlations and on the measurement basis.  In the fully marked eraser example of Sec.~\ref{sec:eraser_example}, the polarization Bloch vector vanishes, so that $p_y=0$, and the entire cohesion budget is stored in the correlation sector.

Equations~\eqref{eq:Pc_full}--\eqref{eq:Pc_sectors} are the main cohesion-accounting result of the paper.  They show that the imaginary-antisymmetric content relevant to phase-sensitive behavior is not exhausted by the reduced path quadrature $V_N=s_y$.  It may also reside in the polarization quadrature $p_y$ or in mixed path--polarization correlations.  These latter contributions vanish from the reduced path Bloch vector but can be accessed by joint path--polarization measurements or converted into path visibility by suitable polarization-erasure operations.  The invariant object is the total scalar cohesion.  The sectoral split in Eq.~\eqref{eq:Pc_sectoral} is a laboratory-basis bookkeeping of where that cohesion is operationally located.

\section{Polarization marking and recoverable coherence}

A pure path--polarization state with one polarization state associated with each slit can be written as
\begin{equation}
|\Psi\rangle=\sqrt{q}\,|1\rangle|\alpha\rangle+
 e^{i\phi}\sqrt{1-q}\,|2\rangle|\beta\rangle,
\label{eq:pure_marker_state}
\end{equation}
where $0\le q\le1$ and $\langle\alpha|\alpha\rangle=\langle\beta|\beta\rangle=1$.  The reduced path coherence is
\begin{equation}
(\rho_{\rm p})_{12}=\sqrt{q(1-q)}\,e^{-i\phi}\langle\beta|\alpha\rangle,
\end{equation}
so that
\begin{equation}
V=2\sqrt{q(1-q)}\,|\langle\beta|\alpha\rangle|,
\qquad
\mathcal P=|2q-1|.
\label{eq:marker_visibility}
\end{equation}
If $|\alpha\rangle$ and $|\beta\rangle$ are orthogonal, the polarization-unresolved visibility vanishes even though the global state is pure.  In that case the lost visibility is not necessarily destroyed; it is stored in path--polarization correlations.  A polarization projection onto a state $|\gamma\rangle$ prepares the conditional path state
\begin{equation}
|\psi_\gamma\rangle\propto
\sqrt q\,\langle\gamma|\alpha\rangle|1\rangle+
 e^{i\phi}\sqrt{1-q}\,\langle\gamma|\beta\rangle|2\rangle,
\label{eq:eraser_conditional}
\end{equation}
whose conditional visibility can be nonzero.  This is the standard quantum-eraser mechanism expressed in the present accounting language: the reduced path vector may lose transverse length while the full path--polarization density matrix retains correlation terms that can be made visible in a different measurement context.

The sectoral quantities above make this distinction explicit.  Decoherence in the path degree of freedom suppresses $s_x$ and $s_y$ and, depending on the channel, may also suppress selected rows of $C$.  Pure polarization marking, by contrast, can reduce $|\bm s_\perp|$ while preserving global purity and redistributing information into $\bm p$ and $C$.  Therefore a decrease of polarization-unresolved visibility is not, by itself, a measure of irreversible coherence loss.  It may be either genuine contraction of the antisymmetric sector, transfer to path--polarization correlations, or a basis-dependent hiding of coherence recoverable by erasure.

\subsection{A fully marked quantum-eraser example}
\label{sec:eraser_example}

Consider the maximally marked state
\begin{equation}
|\Psi_m\rangle=\frac{|1H\rangle+i|2V\rangle}{\sqrt2}.
\label{eq:max_marked_state}
\end{equation}
The reduced path state is maximally mixed,
\begin{equation}
\rho_{\rm p}=\frac12 I,
\qquad
V_A=V_N=\mathcal P=0,
\qquad
\mathcal I=1.
\end{equation}
Thus the polarization-unresolved interferogram has no visibility.  However, the full state is pure.  Its nonzero Bloch--correlation coefficients are
\begin{equation}
C_{xy}=1,
\qquad
C_{yx}=1,
\qquad
C_{zz}=1,
\label{eq:marked_coefficients}
\end{equation}
while $\bm s=\bm p=0$.  Therefore $P_\Delta=1$, and the laboratory-basis cohesion is entirely correlation based:
\begin{equation}
P_c^2=P_{c,{\rm corr}}^{2({\rm glob})}
=\frac12(C_{xy}^2+C_{yx}^2)=1.
\label{eq:marked_cohesion}
\end{equation}
The reduced path measurement has lost all transverse Bloch-vector length, but the complete state still stores phase-sensitive content in the mixed path--polarization cohesion sector.

Now project the polarization onto the diagonal state
\begin{equation}
|D\rangle=\frac{|H\rangle+|V\rangle}{\sqrt2}.
\end{equation}
The successful projection occurs with probability $1/2$ and prepares the conditional path state
\begin{equation}
|\psi_D\rangle=\frac{|1\rangle+i|2\rangle}{\sqrt2}.
\label{eq:conditional_D}
\end{equation}
For this postselected state,
\begin{equation}
V_A=0,
\qquad
V_N=1,
\qquad
\mathcal P=0,
\qquad
P_{c,{\rm p}}^2=1.
\label{eq:erased_path_cohesion}
\end{equation}
The anti-diagonal projection $|A\rangle=(|H\rangle-|V\rangle)/\sqrt2$ gives the opposite quadrature visibility, $V_N=-1$.  This elementary example displays the intended accounting explicitly: before erasure, the reduced path visibility is zero and the cohesion is stored in path--polarization correlations; after conditioning on a polarization-erasing basis, the same phase-sensitive content appears as path cohesion and produces observable fringes in the corresponding postselected subensemble.  No contradiction with reduced complementarity arises, because the reduced identity refers to the polarization-unresolved path state, whereas the eraser uses a different measurement context.

\section{Simple dephasing dynamics}

For completeness, consider a path-dephasing channel acting on the reduced path coherence as
\begin{equation}
(\rho_{\rm p})_{12}(t)=e^{-\Gamma t}(\rho_{\rm p})_{12}(0),
\qquad
\rho_{{\rm p},11}(t)=\rho_{{\rm p},11}(0),
\qquad
\rho_{{\rm p},22}(t)=\rho_{{\rm p},22}(0).
\end{equation}
Then
\begin{equation}
V_A(t)=e^{-\Gamma t}V_A(0),
\qquad
V_N(t)=e^{-\Gamma t}V_N(0),
\qquad
\mathcal P(t)=\mathcal P(0),
\label{eq:dephasing_components}
\end{equation}
and
\begin{equation}
\mathcal I^2(t)=1-e^{-2\Gamma t}\left[V_A^2(0)+V_N^2(0)\right]-\mathcal P^2(0).
\label{eq:I_dynamics}
\end{equation}
The reduced identity is preserved at all times.  In the full $4\times4$ state, however, different physical channels may have the same reduced action and different effects on $\bm p$, $C$, and the cohesion sectors in Eq.~\eqref{eq:Pc_sectors}.  For example, a dephasing channel acting only on the polarization subsystem can leave the reduced path state unchanged while contracting selected correlation-sector cohesion coefficients.  This is another reason to keep the complete path--polarization state whenever polarization acts as a marker, controller, or reservoir of recoverable coherence.

\section{Relation with previous approaches}

The present formulation is consistent with established complementarity relations.  The Greenberger--Yasin and Englert inequalities concern the trade-off between visibility and path information~\cite{Greenberger1988,Englert1996}.  The Jakob--Bergou relation completes this trade-off for pure bipartite path--marker states by adding an entanglement term~\cite{JakobBergou2010}.  Norrman \textit{et al.} developed a more general vector-light treatment in which polarization modulation controls distinguishability and visibility in double-pinhole photon interference~\cite{Norrman2017}.  Maleki emphasized the stereographic geometry underlying coherence and which-path information~\cite{Maleki2019}.  These results already cover the reduced two-level geometry and important polarization-mediated complementarity effects.

The added value of the present approach is narrower but distinct.  It provides an explicit real-basis and IRB-compatible separation of the full path--polarization density matrix into its real symmetric sector and its imaginary antisymmetric cohesion sector, then resolves the latter into path, polarization, and correlation contributions.  It therefore answers a different bookkeeping question: not only how much reduced visibility is observed, but where the corresponding phase-sensitive content resides in the complete state and whether it has been destroyed or merely displaced into degrees of freedom that are ignored by the reduced path measurement.

\section{Conclusions}

The identity $V_A^2+V_N^2+\mathcal P^2+\mathcal I^2=1$ is an exact and useful expression of the reduced path-qubit Bloch geometry, but it should not be regarded as a new complementarity law.  Its role in this paper is to serve as the marginal limit of a full path--polarization accounting.

Keeping the complete $4\times4$ path--polarization density matrix gives the balance
\begin{equation*}
3P_\Delta^2=
V_A^2+V_N^2+\mathcal P^2+P_{\rm pol}^2+\|C\|_F^2,
\end{equation*}
which separates reduced path information, polarization information, and path--polarization correlations.  In a fixed real product basis, the decomposition $\rho=S+i\mathcal T$ further isolates the laboratory representation of the cohesion sector, with
\begin{equation*}
P_c^2=\frac12\left(s_y^2+p_y^2+C_{yx}^2+C_{yz}^2+C_{xy}^2+C_{zy}^2\right).
\end{equation*}
This gives a sector-resolved measure of cohesion in the complete state.  It clarifies why loss of polarization-unresolved visibility can correspond to genuine decoherence, to redistribution into path--polarization correlations, or to recoverable hiding under a quantum-eraser measurement.

This arXiv version is intended as a corrected and more precise formulation of the original preprint.  It establishes the full-state bookkeeping framework and the associated purity and cohesion decompositions.  A more complete journal version will develop the dynamical theory of sectoral cohesion under general path, polarization, and correlated noise channels, and will include explicit tomographic protocols for measuring the antisymmetric coefficients of the full path--polarization state.

\end{document}